\documentclass[12pt]{article}
 \usepackage[english]{babel}
\usepackage{mathtext}
\usepackage[T2A]{fontenc}
\usepackage{epsfig}
\newcommand{\di}{\displaystyle}
\renewcommand{\dh}{\hspace{-2 mm}}
\begin{document}
\renewcommand{\figurename}{Fig.}
\begin{center} {\Large {\bf On the Renyi entropy, Boltzmann Principle, Levy and power-law distributions and Renyi parameter}}\\
  \vspace{.2cm}
\renewcommand{\thefootnote}{\fnsymbol{footnote}}
\vspace{.5cm} \large {\bf A.G.Bashkirov}\footnote{{\it E-mail
address}: abas@idg.chph.ras.ru}\\ \vspace{.25cm} Institute
Dynamics of Geospheres, RAS,\\ Leninskii prosp. 38 (bldg.1),
117334, Moscow, Russia\\
\end{center}
\begin{abstract}
The Renyi entropy with a free Renyi parameter $q$ is the most
justified form of information entropy, and the Tsallis entropy may
be regarded as a linear approximation to the Renyi entropy when
$q\simeq 1$. When $q\to 1$, both entropies go to the
Boltzmann--Shannon entropy. The application of the principle of
maximum of information entropy (MEP) to the Renyi entropy gives
rise to the microcanonical (homogeneous) distribution for an
isolated system. Whatever the value of the Renyi parameter $q$ is,
in this case the Renyi entropy becomes the Boltzmann entropy
$S_B=k_B\ln W$, that provides support for universality of the
Boltzmann's principle of statistical mechanics. For a system being
in contact with a heat bath, the application of MEP to the Renyi
entropy gives rise to Levy distribution (or, $q$-distribution)
accepted as one of the main results of the so-called nonextensive
statistics. The same distribution is derived here for a small
physical system experiencing temperature fluctuations. The
long--range "tail" of the Levy distribution is the power--law
(Zipf-Pareto) distribution with the exponent $s$ expressed via
$q$. The exponent and free Renyi parameter $q$ can be uniquely
determined with the use of a further extension of MEP. Then
typical values of $s$ are found within the range $1.3\div 2$ and
of $q$ within the range $0.25\div 0.5$, in dependence on
parameters of stochastic systems.
\end{abstract}
 \vspace{.1cm}
\vspace{1mm} {\it Keywords:} Renyi entropy, Tsallis entropy,
Escort distribution, Temperature fluctuations.
 \setcounter{footnote}{0}
\section{Introduction} In view of the rapid development of
non-extensive statistics based on Tsallis' information entropy the
problem of choosing the form of information entropy becomes
pressing. The most justified one appears to be one-parameter
family of Renyi's entropies (or simply Renyi entropy). When Renyi
parameter $q$ is equal to unity, Renyi entropy goes to well-known
Boltzmann--Shannon entropy. The Renyi entropy is extensive but its
linearization in the neighbourhood of a point $q\simeq 1$ results
in the non-extensive Tsallis' entropy.

When the principle of maximum of an information entropy (MEP) is
applied to the Renyi entropy of an isolated system, a homogeneous
(microcanonical) distribution $p_i=1/W$ over $W$ micro-states is
derived. In this case, the Renyi entropy becomes the Boltzmann
entropy $S_B=k_B\ln W$ thus afforcing {\it Boltzmann Principle}
from which all thermodynamic properties of extensive and
non-extensive Hamiltonian systems can be deduced.

For a system in contact with a heat bath, the Levy distribution
(or $q$-distribution) $\{p_i\}$ over microscopic states of a
system is derived with the use of MEP applied to the Renyi
entropy. Applying this principle to the Tsallis entropy one would
be forced to go from the original distribution $\{p_i\}$ to an
escort distribution $\{P_i\}=\{p_i^q/\sum_i p_i^q\}$, and use it
further for evaluations of mean values of dynamical variables.
Such going contradicts to main principles of probability
description.

The Levy distribution can be obtained also for small subsystems
experiencing large fluctuations of temperature by way of averaging
of Gibbs distribution over the temperature fluctuations. Besides,
this procedure permits to associate the Renyi parameter $q$ with
physical properties of the subsystem.

A long--range "tail" (for $i\gg 1$) of the Levy distribution can
be approximated by a power--law distribution $p_i\sim x_i^{-s}$,
where the exponent $s$ is expressed in terms of $q$. Then at least
for power--law distributions, the MEP can be expanded to determine
the Renyi parameter $q$ and the exponent $s$, respectively.

\section{Information entropy}

The information entropy, or simply entropy, is the measure of
uncertainty in information in the case of statistical (incomplete)
description of a system using the distribution of probabilities
$p=\{p_i\},\,\,0\leq p_i\leq 1,\,\,i=1,...,W.$ The best known is
the representation of entropy in the Boltzmann--Shannon form
\begin{equation}
S_B=-k_B\sum_i^W p_i\ln p_i.
\end{equation}
In the case when the subscripts $i$ indicate dynamic microstates
in the Gibbs phase space and the distribution ${p_i}$ corresponds
to the macroscopic equilibrium state of the system, the entropy
$S_B$ coincides with the thermodynamic entropy.

It is just this type of entropy was justified by Khinchin [1] and
Shannon [2] in a theorem form based on a system of axioms. Their
axioms were analyzed in [3,4], where it was shown that a unique
determination of entropy in the Boltzmann--Shannon form is
provided by a quite artificial axiom related to a form of
conditional entropy (that is, the entropy of subsystem of a system
being in a prescribed state). A variety of papers on this subject
were analyzed by Uffink [4] who found that the most convincing
appears to be the system of axioms of Shore and Johnson [5]
leading to Renyi's one-parameter family of entropies [6]; for the
distribution $\{p_i\}$ normalized to unity, this family is written
in the form
\begin{equation}
S_R^{(q)}(p)=\frac {k_B}{1-q}\ln
\sum_i^Wp_i^q,\,\,\,\,\,\sum_i^Wp_i =1,
\end{equation}
where $q$ is an arbitrary positive number (it cannot be less than
zero, because $\{p_i\}$ may include zero values). Various
properties of Renyi entropy are discussed, in particular, in the
monographs [6-8]. Among its basic properties we may mention:
positivity ($S_R^q\geq 0$), concavity for $q\leq 1$ and, in
addition, $\lim_{q\rightarrow 1}S_R^q=S_B$.

In the case of $|1-\sum_i p_i^q|\ll 1$ (which, in view of
normalization of the distribution $\{p_i\}$, corresponds to the
condition $|1-q|\ll 1$), one can restrict oneself to the linear
term of logarithm expansion in the expression for $S_R^{(q)}(p)$
over this difference, and $S_R^{(q)}(p)$ changes to
\begin{equation} S_T^{(q)}(p)=-\frac {k_B}{1-q}(1-\sum_i^W p_i^q).
\end{equation}
Such a linearization of Renyi entropy was first suggested by
Havdra and Charvat [9] and Daroczy [10]; at present, this
expression for entropy came to be known as Tsallis' entropy [11].

The logarithm linearization results in the entropy becoming
nonextensive. This property is widely used by Tsallis and by the
international scientific school that has developed around him for
the investigation of diverse nonextensive systems [11-16]. In so
doing, the above-identified restriction $|1-q|\ll 1$ is
disregarded, as a rule. In our opinion, the attempt by Abe [16] at
independent validation of this form of entropy appears
unconvincing, because it is based on the axiomatic introduction of
such a form for conditional entropy which uniquely provides for
obtaining Tsallis' entropy.

\section{Extremality of entropy}

According to MEP, in the case of statistical (incomplete, from the
dynamic standpoint) description of the system, its distribution
function must provide for correct values of those few average
quantities which appear in the statistical description; otherwise,
it must be as undetermined as possible. Such an approach in
application to equilibrium thermodynamic systems (isolated or
weakly interacting with the thermostat) has long been used to
construct equilibrium statistical thermodynamics. However, it was
only after studies by Jeynes [17] that it came to be firmly
established as a principle validating (at least, on the physical
level of rigor) the use of Gibbs ensembles in statistical
description of thermodynamic systems. The information entropy is
traditionally taken to mean the Boltzmann--Shannon entropy. Here,
MEP will be applied to the Renyi entropy.

\subsection{Microcanonical distribution. Boltzmann principle.}
We are interested firstly in the distribution of probabilities
$\{p_i\}$, providing for the extremality of information entropy
with an additional condition of normalization of $p_i$.

Then, the distribution $\{p_i\}$ must be determined from the
extremum of the functional
 \begin{equation} L^{M}_R(p )=\frac
1{1-q}\,\ln \sum_i^W\,p^{q } _i - \Phi \sum_i^W\,p _i,
\end{equation}
where $ \Phi $ is the Lagrange multiplier dependent on $q$.

We equate its functional derivative to zero to derive
\begin{equation}
\frac{\delta L^{M}_R(p )}{\delta p _i}=\frac q{1-q}\,\frac
{p_i^{q-1}}{\sum_j p_j^{q}} - \Phi =0.
\end{equation}
 Multiplying this equation by $p_i$ and
summing up over $i$, with account of normalization condition
$\sum_i p_i=1$ we get $\Phi =\frac q{1-q}$. Then, it follows from
equation (4) that
$$
p _i=\left(\sum_j^W\,p^{q} _j\right)^{\frac{1}{q-1}}. $$
Using once more the condition $\sum_i p_i=1$ we get $$
\left(\sum_i^W\,p_i^q\right)^{\frac 1{q-1}} =\frac 1{W} $$ and,
finally,
\begin{equation}
p_i =\frac 1{W}\,\,.
\end{equation}
Thus, we have obtained equally probable distribution ${p_i}$
corresponding to the microcanonical Gibbs ensemble of statistical
mechanics. The Renyi entropy for this case takes a form
\begin{equation}
S^{M}_R=\frac {k_B}{1-q}\,\ln W^{1-q } = k_B\,\ln W.
\end{equation}
This expression does not depend on the Renyi parameter and  at any
$q$ coincides exactly with the Boltzmann entropy definition
$S_{B}=k_B\ln W$ called by Einstein as {\it Boltzmann's Principle}
from which all equilibrium statistical thermodynamics both
extensive and non-extensive Hamiltonian systems can be deduced
(see, e.g. [18]). It was proposed [18] to take  Boltzmann
Principle as the axiomatic assumption. Here we obtained it as a
consequence of the axiomatics of Shore and Johnson [5] led us to
Renyi entropy and then to Boltzmann Principle.

Thus, the Renyi entropy gives new physical insight into the
Boltzmann Principle. The Tsallis entropy exhibits this mandatory
property only when $q\to 1$.

\subsection{Canonical Levy distribution.}
Next we will look for the distribution of probabilities $\{p_i\}$,
providing for the extremality of information entropy with an
additional conditions which consist in preassigning the average
value $\bar H=\langle H\rangle_p\equiv\sum_i H_i p _i$ of the
random quantity $H_i$ and the requirement of normalization of
$p_i$.

Then, the distribution $\{p_i\}$ must be determined from the
extremum of the functional
 \begin{equation} L_R(p )=\frac
1{1-q}\,\ln \sum_i^W\,p^{q } _i - \beta_0 q \,\sum_i^W H_i p _i -
\Phi \sum_i^W\,p _i,
\end{equation}
where $\beta_0 q$ and $ \Phi $ are Lagrange multipliers dependent
on $q$. Note that, in the $q\to 1$ limit it changes to well-known
functional
 \begin{equation}
L_G(p )=- \sum_i^W\,p_i\ln p_i - \beta_0  \,\sum_i^W H_i p _i -
\Phi_0 \sum_i^W\,p _i;
\end{equation}
its extremum is ensured by the Gibbs canonical distribution, in
which $\beta_0 =1/k_BT_0$ and $\Phi_0 $ is the free energy.

We equate a functional derivative of $L_R(p )$ to zero to derive
 \begin{equation}
\frac{\delta L_R(p )}{\delta p _i}=\frac q{1-q}\,\frac
{p_i^{q-1}}{\sum_j p_j^{q}}-\beta_0 qH_i - \Phi =0.
\end{equation}
Multiplying this equation by $p_i$ and summing up over $i$, with
account of normalization condition $\sum_i p_i=1$ we get $\Phi
=\frac q{1-q}- \beta_0 q \bar H$. Then, it follows from equation
(10) that
$$
p _i=\left(\sum_j^W\,p^{q} _j\, (1+\beta_0(1-q)(H_i-\bar
H))\right)^{\frac{1}{q-1}}. $$
Using once more the condition $\sum_i p_i=1$ we get $$
\sum_j^W\,p_j^q =\left(\sum_i^W (1+\beta_0(1-q)(H_i-\bar
H)^{\frac{1}{q-1}}\right)^{-(q-1)}$$ and, finally,
\begin{equation}
p_i =\frac {\left(1+\beta_0(1-q)(H_i-\bar
H)\right)^{\frac{1}{q-1}}} {\sum_i\left(1+\beta_0(1-q)(H_i-\bar
H)\right)^{\frac{1}{q-1}}}.
\end{equation}
Such distribution is known now as Levy distribution or
$q$--distribution. At $q\to 1$ the distribution $\{p_i\}$ becomes
the Gibbs canonical  distribution in which the constant
$\beta_0=1/k_BT_0$ is the reciprocal of the temperature.

If the Tsallis entropy was used instead of the Renyi entropy the
Levy distribution was derived [11] in the form
\begin{equation}
p^{Ts}_i =\frac {\left(1+\beta_0(1-q)H_i\right)^{\frac{1}{q-1}}}
{\sum_i\left(1+\beta_0(1-q)H_i\right)^{\frac{1}{q-1}}},
\end{equation}
but the starting functional was forced to be taken as
\begin{equation} L^{Ts}(p )=-\frac
1{1-q}\,\left(1- \sum_i^W\,p^{q } _i\right) - \Phi\beta_0 (q-1)
\,\sum_i^W H_i p _i + \Phi \sum_i^W\,p _i.
\end{equation}
Here, the question arises about forms of Lagrange multipliers
$\Phi\beta_0(q-1)$ and $(+\Phi)$, but the main problem is that the
functional $L^{Ts}(p )$ does not pass to the functional (9) when
$q\to 1$ as the second term in (13) vanishes.

It seems reasonable to suppose that just this difficulty causes to
use the escort distribution  $P_i=p_i^q/\sum_ip_i^q$ in
nonextensive thermodynamics [12,13,15]. The consistency of the
transition to escort distribution is partly justified by the
condition conservation of a preassigned average value of the
energy $\bar H=\langle H\rangle_{es}\equiv\sum_i H_i\,P_i$,
however other average values are to be calculated with the use of
the same escort distribution also, that contradicts to the main
principles of probability description.

It should be noted also that escort distributions were introduced
[7] as a tool to scan the structure of an original distributions
$\{p_i\}$. Indeed, at $q>1$, the importance of $p_i$ with the
maximal values increases, and at $q<1$, of $p_i$ with minimal
values. In view of this, it is evident that use of escort
distributions in statistical thermodynamics does not lead to true
average values of dynamical variables.

\section{Small subsystem with fluctuating temperature}

To clear up a physical sense of the Levy distribution (11) and the
Renyi parameter $q$ we use here an approach proposed by Wilk and
Wlodarchuk [19].

We will treat a subsystem which is a minor part of a large
equilibrium system and experiences thermal fluctuations of both
energy and temperature. This is a radical difference of the
suggested approach from the Gibbs approach traditionally employed
in statistical physics, in which temperature is preassigned by a
constant characterizing the thermostat.

In equilibrium theory of thermodynamic fluctuations the mean
square temperature fluctuation is estimated as $\delta
T/T_0=(k_B/C)^{1/2}$ where $C$ is the heat capacity of the
subsystem. More detailed analysis [20], based on the
Landau--Lifshits theory of hydrodynamic fluctuations, gives the
next nonlinear stochastic Langevin equation for fluctuating
temperature
\begin{equation}
 \frac {dT(t)}{d t}=-\frac 1{\tau}(T(t)- T_0) - \frac 1{C}T(t)
\xi(t),
\end{equation}
where $\tau =C/(A\kappa)$, $A$ is the surface area of the
subsystem, $\kappa $ is the heat transfer coefficient, and $T_0$
is the average temperature of the subsystem. A random function of
time $\xi(t)$ satisfies the relation
 \begin{equation}
\langle \xi(t)\rangle =0,\,\,\,\langle \xi(t)\xi(t')\rangle
=2k_BA\kappa \delta (t-t').
\end{equation}

Corresponding to the derived stochastic Langevin equation with
$\,\delta$-correlated noise is the Fokker--Planck equation for the
temperature distribution function $f(T,t)$
\begin{equation}
\frac {\partial f(T,t)}{\partial t}= -\frac {\partial }{\partial
T}W_1(T)f(T,t)+\frac 1{2}  \frac {\partial^2 }{\partial
T^2}W_2(T)f(T,t).
\end{equation}
The coefficients $W_1(T)$ and $W_2(T)$ of this equation are
expressed in terms of the first $\langle T(t)-T(t+\tau)\rangle $
and second $\langle (T(t)-T(t+\tau))^2\rangle $ conditional
moments of stochastic equation Langevin. A steady-state solution
[20,21] to equation (16) is
\begin{equation}
f(z)=\frac {1}{\Gamma (\gamma)}z^{\gamma -1}e^{- z}.
\end{equation}
where the dimensionless constant $\gamma=C /k_B$ is introduced,
and $z=\gamma T_0/T$. Therefore, the thus derived distribution
function of the inverse temperature of the subsystem in the
dimensionless form is the gamma distribution.

If the mean energy of the singled-out volume $\bar H = CT_0$ is
introduced, this expression may be rewritten as
\begin{equation}
f(\beta)=\frac {(\gamma \beta/\beta_0)^\gamma}{\beta\Gamma
(\gamma)}\,e^{-\di\beta\bar H}.
\end{equation}
By its form, this distribution is close to the Gibbs distribution;
however, unlike the latter, it accounts for the temperature
fluctuation of the subsystem with the preassigned mean energy
$\bar H $.

In order to describe a subsystem in contact with a large thermally
equilibrium system (thermostat), the Gibbs canonical distribution
is used in statistical physics (here and below, the factor $G_i$
allowing for number of states of energy $H_i$ is omitted for
brevity):
\begin{equation}
\rho_i=Q^{-1}e^{-\beta H_i},
\end{equation}
where $H_i$ is the energy of the subsystem (the subscript $i$ may
indicate the number of discrete energy level or totality of the
values of coordinates and momenta of molecules of the subsystem),
and $Q$ is the partition function. In so doing, the inverse
temperature $\beta=1/k_BT$ is taken to be known preassigned
quantity.

As was demonstrated above, the temperature may fluctuate. In view
of this, the question arises as to how the Gibbs distribution is
modified under the effect of temperature fluctuations. The answer
to this question may be obtained by the way of averaging the Gibbs
distribution (19) with the gamma-distribution for temperature $T$
(or $\beta$).

For further treatment, $\rho_i$ may be conveniently represented in
an equivalent form,
\begin{equation}
\rho_i=Q^{-1}e^{-\beta \Delta H_i},\,\,\,Q=\sum_i e^{-\beta \Delta
H_i },\,\,\,\Delta H_i =H_i-\bar H
\end{equation}
where the symbol $\sum_i$ may indicate both the summation and
integration over a totality of the values of coordinates and
momenta.

Using the mean value theorem we represent the Gibbs distribution
averaged over $\beta $ in the form
\begin{equation}
\bar \rho_i=\int_0^\infty d\beta f(\beta)\rho_i=\frac
1{Q^*}\int_0^\infty d\beta f(\beta)e^{-\beta \Delta H_i},
\end{equation}
where $Q^*$ lies in the range of possible variation of $Q(\beta )$
from $Q(0)$ to $Q(\infty)$. From the conditions of normalization
to unity of the distributions $f(\beta)$ and $\rho$, we have
\begin{equation}
\frac 1{Q^*}\sum_i\int_0^\infty d\beta f(\beta)e^{-\beta \Delta
H_i}=1
\end{equation}
whence we find
\begin{equation}
Q^*=\sum_i\int_0^\infty d\beta f(\beta)e^{-\beta \Delta H_i}.
\end{equation}
Therefore, it is sufficient to calculate only the average value of
the exponent,
\begin{equation}
\int_0^\infty d\beta f(\beta)e^{-\beta \Delta H_i}=\frac {(\gamma
k_BT_0)^\gamma}{\Gamma (\gamma)}\beta^{\gamma -1}\int_0^\infty
d\beta \beta^{\gamma -1}e^{-\beta(\gamma k_BT_0 +\Delta
H_i)}=\Bigl(1+\frac {\beta_0}{\gamma}\Delta H_i\Bigr)^{-\gamma}.
\end{equation}
Finally, the averaged Gibbs distribution takes the form
\begin{equation}
\bar\rho_i=\frac {\left(1+\frac {\beta_0}{\gamma}(H_i-\bar H
)\right)^{-\gamma}}{\sum_i\left(1+\frac {\beta_0}{\gamma}(H_i-\bar
H )\right)^{-\gamma}}.
\end{equation}
In the $\gamma\to\infty$ limit corresponding to a high heat
capacity of the singled-out subsystem, $\bar\rho_i$ goes to
$\rho_i$.

Resulted equation for the modified Gibbs distribution is similar
to equation (11) for $p_i$ in its structure. To identify
$\bar\rho_i$ with $p_i$ it is enough to present $\gamma$ as
$\gamma=(1-q)^{-1}$, then equation (25) takes the form
\begin{equation}
\bar\rho_i=\frac {\left(1+\beta_0 (1-q)(H_i-\bar H
)\right)^{-\frac 1{1-q}}}{\sum_i\left(1+\beta_0 (1-q)(H_i-\bar H
)\right)^{-\frac 1{1-q}}}.
\end{equation}
The full identity of this expression with the probability $p_i$
(11) ensuring the extremality of Renyi entropy enables one to take
a new view of the physical meaning of the Renyi entropy and
parameter
\begin{equation}
q=1-\frac 1{\gamma}=\frac {C-k_B}{C}.
\end{equation}
So, the Renyi parameter differs significantly from unity only in
the case where the heat capacity of the singled out system is of
the same order of magnitude as the Boltzmann constant $k_B$. The
thermodynamics of such systems must be constructed on the basis of
Renyi entropy and Boltzmann Principle or the Levy distribution
function $\{\bar \rho_i\}$ or $\{p_i\}$ for systems in contact
with a heat bath. We shall emphasize once again that here this
function was obtained without invoking any additional
considerations as to the nonextensiveness of the systems being
treated.

When $i$-th state of the system is determined by a value $H_i$
(indexed in increasing order of magnitude, so that $H_{max}=H_W$),
and $x_i=H_i/H_W$ then for $q\neq 1$ and sufficiently significant
fluctuations of $x$ exceeding the minimal value
\begin{equation} x_{min}\gg
\left|\bar x -\frac 1{\beta_0 H_W(1-q)}\right|,
\end{equation}
expression (6) transforms to the power distribution
\begin{equation}
p_i\sim x_i^{-s},\,\,\,\,s=\frac{1}{1-q}.
\end{equation}
So, we derived the Zipf--Pareto power--law distribution  ensuring
the extremality of Renyi entropy at $q\neq 1$. In this aspect it
can be said that the power--law distribution is inherent to the
Renyi entropy as much as Boltzmann or Gibbs distributions are
inherent to the Boltzmann--Shannon entropy. As the exponent $s$ is
expressed (29) in terms of the Renyi parameter $q$, the evident
requirement $s>0$ produces a new constraint $q< 1$ that coincides
with the condition of concavity of the Renyi entropy\footnote{When
one tries to derive the power--law distribution on the base of the
escort distribution he obtains the exponent $s_{es}=q/(q-1)$ that
produces a constraint $q>1$. For such $q$ the Renyi entropy is nor
pure concave.}. Otherwise the Renyi parameter remains arbitrary,
so while it is varying from $q\ll 1$ to $q\simeq 1$ the exponent
$s$ varies from $1$ to $+\infty$. Hence it is evident that an
additional physical concept should be invoked to specify $q$
uniquely.

\section{Extension of MEP to determine Renyi parameter}

As a rule, the exponent $s$ of power--law distributions for
stochastic systems of different nature lies in the narrow range of
values between 1 and 2. It follows from it that a corresponding
value of the Renyi parameter $q$ connected with $s$ by equation
(29) lies too in a narrow range.

In this connection, it is reasonable to assume an existence of
some variational principle providing realization of value of the
Renyi parameter just in this range. Inasmuch as the Levy
distribution possessing the power--law "tail" is inherent in the
Renyi entropy, it seems to be reasonable to look for extremum on
$q$ of the difference
\begin{equation}
\Delta S = S^{(q)}_R(\{p_i\}) - S_B(\{p_i\})
\end{equation}
where $\{p_i\}$ is the Levy distribution (11).

When the subscript $|\Delta p_i|\equiv |p_{i+1}-p_i|\ll p_i$ for
all $i$, we can pass to a continuous picture. Then the probability
$p_i$ is replaced by $p(x_i)\Delta x_i$, where $p(x)$ is the
probability density and $\Delta x_i=x_{i+1}-x_i$. For simplicity,
we assume that all $\Delta x_i$ are of the same value ($\Delta
x_i=\Delta x$ for all $i$), then we get
\begin{equation}
S^{(q)}_R(\{p_i\})=\frac 1{1-q}\ln\sum_i^W p^q(x_i)(\Delta
x)^q=-\ln\Delta x +S^{(q)}_{RD}(p(x)),
\end{equation}
\begin{equation}
S_B(\{p_i\})=-\sum_i^W p(x_i)(\Delta x)\ln [p(x_i)(\Delta
x)]=-\ln\Delta x +S_{BD}(p(x))
\end{equation}
where
\begin{equation}
S^{(q)}_{RD}(p(x))=\frac 1{1-q}\ln\int_{x_1}^1
p^q(x)dx,\,\,\,S_{BD}(p(x))=-\int_{x_1}^1 p(x)\ln p(x)dx.
\end{equation}
It is suggested that the probability density and differential
$\Delta x$ satisfy all requirements which are necessary when
passing from Darboux sums to integrals in equations (31) and (32).

It is conventional that $S_{BD}$ is referred to as the
differential Boltzmann entropy, so we call $S^q_{RD}$ differential
Renyi entropy. The original Boltzmann and Renyi entropies are
positive defined values, but differential ones are not positive
defined due to smallness of $\Delta x$. Nevertheless, their
difference is the same as the difference of original entropies, so
that
\begin{equation}
\Delta S = S^{(q)}_{RD}(p(x)) - S_{BD}(p(x)).
\end{equation}

\begin{figure}
\begin{minipage}{.50\linewidth}
 \centering\epsfig{figure=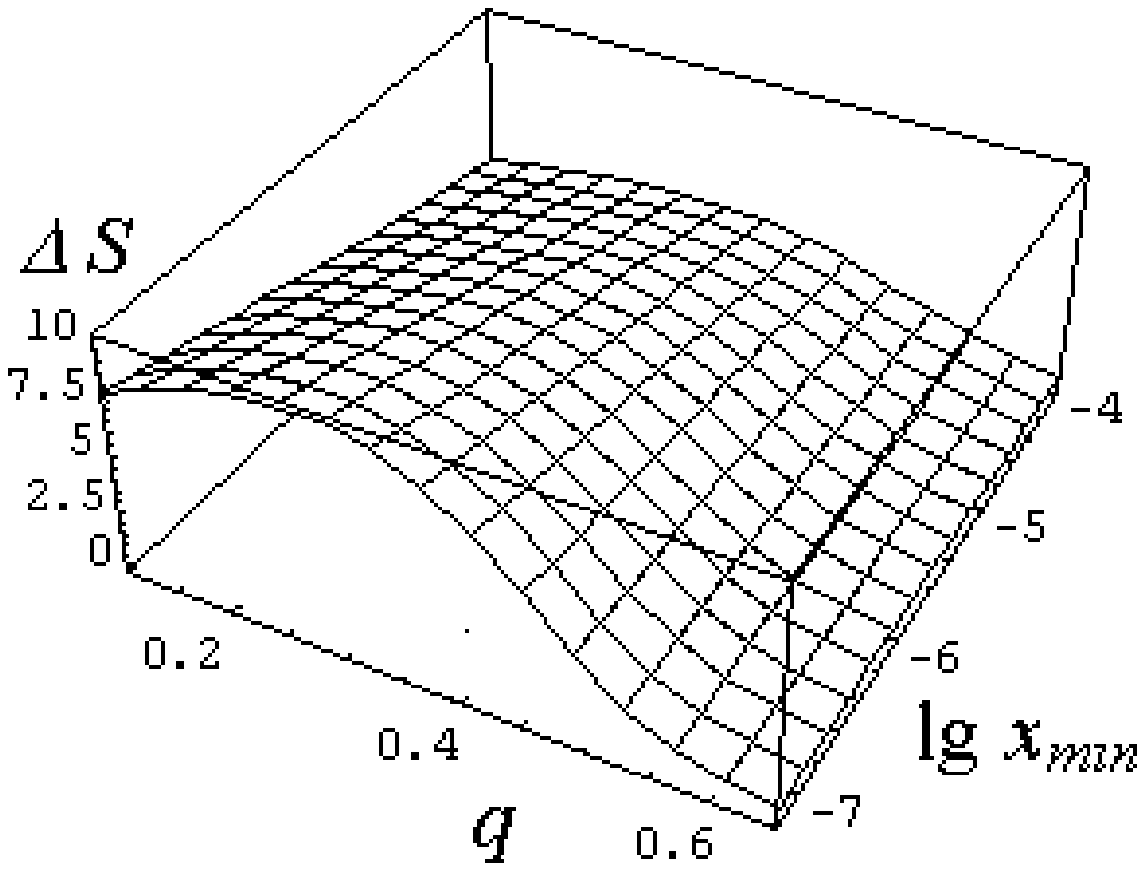, height=5cm}
 \end{minipage}
\begin{minipage}{.55\linewidth}
 \centering\epsfig{figure=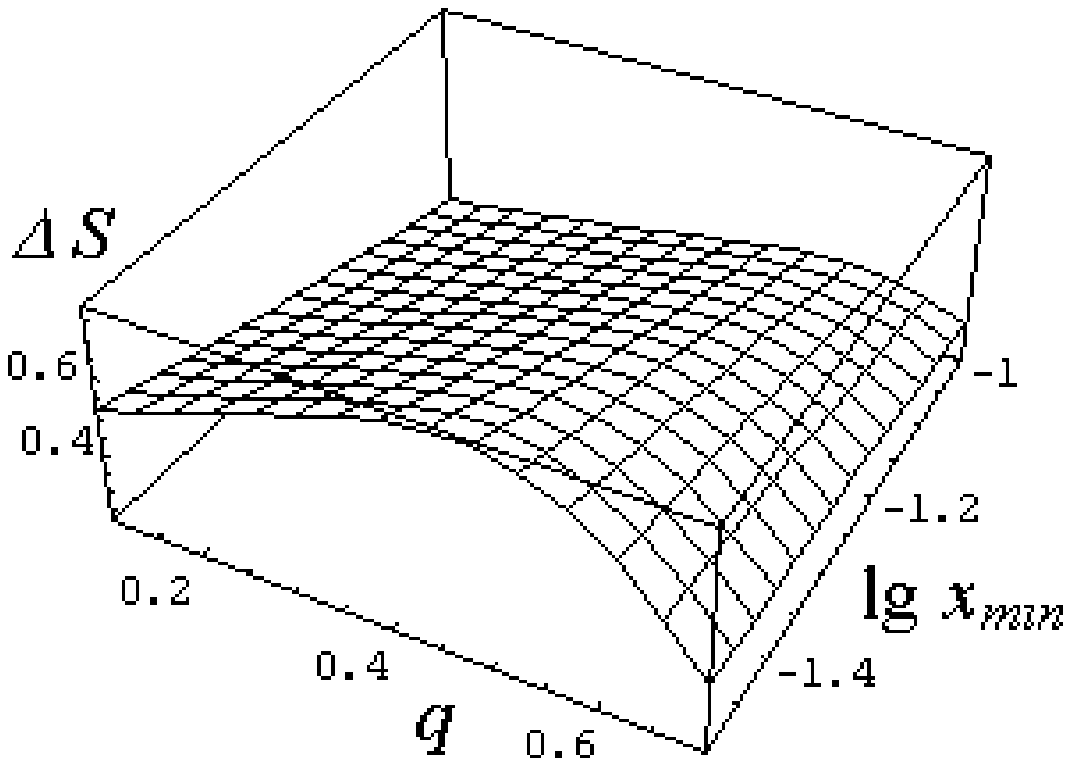, height=5cm}
  \end{minipage}
\caption{The difference $\Delta S=S^{(q)}_{R}-S_{B}$ for
power--law distribution $p\sim x^{-1/1-q}$ as a function of $q$
and $x_{min}$.}
\end{figure}

An investigation of this function of $q$ for the presence of its
extremum is only possible for a concrete system when all
parameters, $\beta_0,\bar H $ of the Levy distribution are known.
However, if a main contributions into the integrals of
differential entropies (34) are provided by the range of $x$
values from $x_{min}$ to $x_{max}=1$, we can use the power--law
distribution density instead of Levy distribution density
\begin{equation}
p(x|s)=B^{-1}x^{-s},\,\,\,B=\int_{x_{min}}^1 x^{-s}\,dx=\frac
1{1-s}(1-x_{min}^{1-s}).
\end{equation}
Then
\begin{eqnarray}
\Delta S(q)\dh &=&\dh S^{(q)}_{RD}(p(x|s))-S_{BD}(p(x|s))
\nonumber
\\ \dh &=&\dh -\frac {B x_{min}^{1-s}}{(s-1)^2}\big[(1 - x_{min}^{s-1})s +
x_{min}^{s-1} (s-1)\ln B \nonumber \\   \dh &+& \dh  (1 - s)\ln (B
x_{min}^{-s})\big]+\frac 1{(1-q)}\ln\left(\frac {B^q(1 -
x_{min}^{1 - q s}}{1 - q s}\right).
\end{eqnarray}
This function of $q$ depends on the parameter $x_{min}$ only. Its
three-dimensional plot is illustrated in Fig. 1. Well defined
maximum with respect to $q$ is seen at it and its position $q^*$
depends slightly on $x_{min}$. Substituting $q=q^*(x_{min})$ into
equation (29), we get $s^*=1/(1-q^*(r))$, which is illustrated in
Fig. 2. It is seen that values of $s^*(x_{min})$ are in the range
from 1.1 to 2 depending on $x_{min}$.
\begin{figure}[t]
 \centering\epsfig{figure=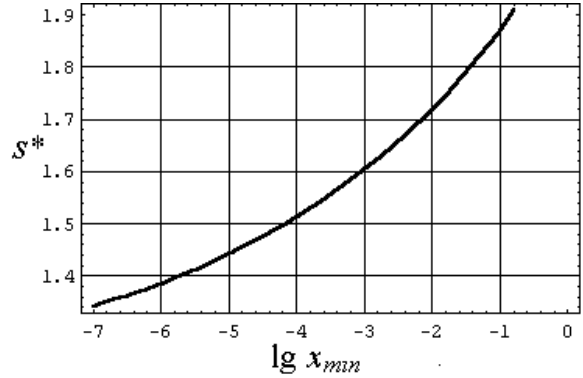, height=5cm}
\caption{The exponent of the power--law distribution
$s^*=1/(1-q^*)$ as a function of $x_{min}$.}
\end{figure}

Distributions of this kind were found for an extremely wide class
of natural and social phenomena. As examples, we can mention the
distributions of magnitudes of earthcrackings (Gutenberg--Richter
law), vortices over their energies in turbulent flows, and also in
the science of sciences [22], economics (Zipf--Pareto distribution
law for citizens over their incoming or enterprizes over number of
collaborators, or banks over their capitals), geography
(distribution law for countries or cities over their citizens)
etc.

In these examples, we deal with parameters being analogs of
energy. In opposite cases, when we are interested in the
power--law distribution over other parameters, say $l$, we should
recalculate the exponent. Really, if $H\sim l^r$ then
$p(x|s)dx\sim H^sdH\sim l^{-rs+r-1}dl$, that is, the modified
exponent is $s_l=rs-r+1$. As an example, for the distribution of
fragments over their masses in impact fragmentation of solids, we
have [23] $H\sim m^{2/3}$ and $x_{min}\simeq 10^{-5}$, then, from
the Fig. 2, we find $s\simeq 1.5$ whence $s_m\simeq 1.3$, that is
in agreement with the experimental results [24] and theoretical
estimations [23].
\subsection*{Acknowledgements} I acknowledge
useful discussions with A.V. Vityazev.
\subsection*{References} [1] A.Ya. Khinchin, Mathematical
Foundations of Information Theory.

Dover Publ., N.-Y., 1957\\[0pt] [2] C.E. Shannon, Bell Syst.
Techn.Journ. {\bf 27} 379 (1948)\\[0pt][3] A.G. Bashkirov, A.V.
Vityazev, Physica A {\bf 277}, 136 (2000).\\[0pt] [4] J. Uffink,
Studies in Hist. and Philos. of Mod. Phys. {\bf 26B}, 223
(1995).\\[0pt] [5] J.E. Shore, R.W. Johnson, IEEE Trans. Inform.
Theory {\bf IT-26}, 26

(1980); {\bf IT-27}, 472 (1981); {\bf IT-29}, 942 (1983).\\[0pt]
[6]  A. Renyi, Probability theory, North-Holland, Amsterdam
(1970).\\[0pt] [7]  C. Beck, F.  Schl\"ogl, Thermodynamics of
Chaotic Systems, Cambridge

Univ. Press, Cambridge (1993).\\[0pt] [8] Yu.L. Klimontovich,
Statistical Theory of Open Systems. Kluwer

Academic Publishers. Dordrecht, 1994\\[0pt] [9] J.S. Havrdra, F.
Charvat, Kybernatica {\bf 3}, 30 (1967)\\ [0pt] [10] Z. Daroczy,
Information and Control {\bf 16}, 36 (1970)\\[0pt] [11] C.
Tsallis, J.Stat.Phys. {\bf 52}, 479 (1988)\\[0pt] [12] E.M.F.
Curado, C. Tsallis, J.Phys. A:Math.Gen. {\bf 24}, L69 (1991)\\
[0pt] [13] M. Casas, S. Martinez, F. Pennini, A. Plastino, Physica
A {\bf 305}, 41 (2002)\\[0pt] [14] A.R. Plastino, A.Plastino,
Physica A {\bf 222}, 347 (1995)\\[0pt] [15] C. Tsallis, R.S.
Mendes, A.R. Plastino, Physica A {\bf 261}, 534 (1998).\\ [0pt]
[16] S. Abe, Phys.Lett. A {\bf 271}, 74 (2000).\\ [0pt] [17] E.T.
Jaynes, Phys.Rev. {\bf 106}, 620; {\bf 107}, 171 (1957)\\[0pt]
[18] D.H.E. Gross, Physica A {\bf 305}, 89 (2002)\\[0pt] [19] G.
Wilk, Z. Wlodarczyk, Phys.Rev.Lett. {\bf 84}, 2770 (2000).\\[0pt]
[20] A.G. Bashkirov, A.D. Sukhanov, J. Exp. Theor. Phys. {\bf 95},
440 (2002)\\ [0pt] [21] S. Akhmanov, Yu.E. D'jakov, A.S. Chirkin,
Introduction to

Statistical Radiophysics and Optics. Nauka, Moscow, 1981\\ [0pt]
[22] D. Price, Little Science, Big Science. N.-Y. Columbia Univ.
1963\\  [0pt] [23] A.G. Bashkirov, A.V. Vityazev, Planet. Space
Sci. v.44, 909--915 (1996) \\[0pt] [24] A. Fujiwara, G. Kamimoto
and A. Tsukamoto,  Icarus, {\bf 31,} 277 (1977)\\ [0pt]
\end{document}